\documentclass[12pt]{spieman}  

\usepackage{amsmath,amsfonts,amssymb}
\usepackage{graphicx}
\usepackage{setspace}
\usepackage{tocloft}
\usepackage{lineno}
\usepackage{here}
\usepackage{color}

\title{X-ray imaging polarimetry with a 2.5-$\mathrm{\mu}$m pixel CMOS sensor for visible light at room temperature}

\author[a]{Kazunori~Asakura}
\author[a,b,c]{Kiyoshi~Hayashida}
\author[a]{Takashi~Hanasaka}
\author[a]{Tomoki~Kawabata}
\author[a]{Tomokage~Yoneyama}
\author[a]{Koki~Okazaki}
\author[a]{Shuntaro~Ide}
\author[a,b]{Hirofumi~Noda}
\author[a,b]{Hironori~Matsumoto}
\author[a]{Hiroshi~Tsunemi}
\author[d]{Hisamitsu~Awaki}
\author[e]{Hiroshi~Nakajima}

\affil[a]{Osaka University, Department of Earth and Space Science, 1-1 Machikaneyama-cho, Toyonaka, Osaka, Japan}
\affil[b]{Osaka University, Project Research Center for Fundamental Sciences, 1-1 Machikaneyama-cho, Toyonaka, Osaka, Japan}
\affil[c]{Institute of Space and Astronautical Science, Japan Aerospace Exploration Agency, 3-1-1 Yoshino-dai, Chuo-ku, Sagamihara, Kanagawa, Japan}
\affil[d]{Ehime University, Department of Physics, 2-5 Bunkyo-cho, Matsuyama, Ehime, Japan}
\affil[e]{Kanto Gakuin University, College of Science and Engineering, 1-50-1 Mutsuura Higashi, Kanazawa-ku, Yokohama, Kanagawa, Japan}

\cftpagenumbersoff{figure}
\cftpagenumbersoff{table}
\begin{document}
\maketitle

\begin{abstract} 

X-ray polarimetry in astronomy has not been exploited well, despite its importance. The recent innovation of instruments is changing this situation. We focus on a complementary MOS (CMOS) pixel detector with small pixel size and employ it as an x-ray photoelectron tracking polarimeter. The CMOS detector we employ is developed by GPixel Inc., and has a pixel size of 2.5\,$\mathrm{\mu}$m $\times$ 2.5 $\mathrm{\mu}$m. Although it is designed for visible light, we succeed in detecting x-ray photons with an energy resolution of 176\,eV (FWHM) at 5.9\,keV at room temperature and the atmospheric condition. We measure the x-ray detection efficiency and polarimetry sensitivity by irradiating polarized monochromatic x-rays at BL20B2 in SPring-8, the synchrotron radiation facility in Japan. We obtain modulation factors of 7.63\% $\pm$ 0.07\% and 15.5\% $\pm$ 0.4\% at 12.4\,keV and 24.8\,keV, respectively. It demonstrates that this sensor can be used as an x-ray imaging spectrometer and polarimeter with the highest spatial resolution ever tested.


\end{abstract}


{\noindent \footnotesize\textbf{*}Kazunori~Asakura, asakura$\_$k@ess.sci.osaka-u.ac.jp}


\section{Introduction}
\label{sec:intro}  

Polarimetry is a powerful diagnostic tool to measure magnetic fields and scattering process of celestial objects. In fact, polarimetry has been widely employed for various targets and purposes in radio-to-ultraviolet astronomy. On the other hand, polarimetry in x-ray and $\gamma$-ray astronomy had long been an unexploited field after the first detection of polarization of the Crab Nebula with a sounding rocket instrument\cite{Novick1972} and succeeding satellite observations\cite{Weisskopf1978}. Polarimetry in hard x-ray and soft $\gamma$-ray astronomy has been markedly activated in these years, starting from the detection of polarization in the Crab Nebula with instruments on-board INTEGRAL\cite{Forot2008,Dean2008}. Hard x-ray to soft $\gamma$-ray polarimetry of the Crab Nebula has succeeded in other instruments, e.g., the CZT Imager on ASTROSAT\cite{Vadawale2018}, PoGO+\cite{Chauvin2017, Chauvin2018}, and the Soft Gamma-ray Detector on Hitomi\cite{Hitomi2018}. Polarization of Cyg X-1 has also been measured with some of these instruments. Gamma-ray bursts are also the targets of extensive observations of their polarization, with GAP on IKAROS\cite{Yonetoku2011}, the CZT Imager on ASTROSAT\cite{Chattopadhyay2017}, and POLAR on-board Tiangong-2\cite{Zhang2019}. The imaging x-ray polarimetry explorer (IXPE)\cite{Weisskopf2016}, a small satellite mission dedicated to soft x-ray polarimetry, is scheduled to be launched in 2021.

Various polarimetry instruments have been developed in x-ray and $\gamma$-ray astronomy, as reviewed by Weisskopf et al.\cite{Weisskopf2010}  They can be classified into three main categories by their working principle: Bragg reflection, Compton/Thomson scattering,  and photoelecton tracking. The first two were employed in real observations, whereas observation with the last one will be achieved with IXPE. Photoelectron tracking polarimetry utilizes the anisotropic distribution of the photoelectron emission in photoelectric absorption of x-rays. The anisotropy is most enhanced in the K-shell photoelectric absorption; its differential cross section is described as
\begin{align}
\label{eq1}
\frac{d{\sigma}}{d{\Omega}}{\propto}\frac{{\sin{\theta}^2}{\cos{\phi}}^2}{(1-{\beta}\cos{\theta})^4},
\end{align}
where $\theta$ and $\phi$ are the polar and azimuth angles of the emitted photoelectron with respect to the x-ray polarization vector, and $\beta$ is the speed of the photoelectron normalized by the speed of light. IXPE employed micropixel gas detectors with gas electron multiplier foils. Similar format detectors were developed by various groups\cite{Sakurai1996, Tanimori1999, Black2003}. Photoelectron tracking polarimeters can also be realized with solid-state detectors. The first demonstration with a CCD was presented by Tsunemi et al.\cite{Tsunemi1992} In that experiment, a CCD with a pixel size of 12\,$\mathrm{\mu}$m was employed. A CCD with a smaller pixel size of 6.8\,$\mathrm{\mu}$m was employed in the following experiment conducted by Buschhorn et al.\cite{Buschhorn1994} Polarimetric performance can be optimized with data reduction.\cite{Hayashida1999} Although x-ray CCDs were utilized in most of the x-ray astronomical satellites since ASCA, no positive detection of polarization has been reported. This is mainly due to the pixel size of the x-ray CCDs used in space, ranging from 24\,$\mathrm{\mu}$m (Chandra ACIS and Suzaku XIS) to  150\,$\mathrm{\mu}$m (XMM-Newton EPIC-pn), which is too large to measure photoelectron tracks.

Comparing the two classes of photoelectron tracking polarimeters, gas detectors and solid-state detectors, the former has higher polarimetry sensitivity. This comes from the difference in the range of electrons; the range is longer in gas than in a solid. However, since the imaging and spectroscopic capability of solid-state detectors is superior to that of gas detectors, there can be a wide range of application fields in solid-state detectors if they have enough polarimetric capability.  Such instruments will be suitable not only for a focal plane detector with an x-ray mirror but also for the newly proposed x-ray interferometer without a mirror\cite{Hayashida2016, Hayashida2018}.

We focus on complementary MOS (CMOS) pixel sensors designed for visible light with a pixel size of several micrometers, which are developed for numerous commercial applications. The noise performance of the latest CMOS pixel sensors is as good as or better than that of CCDs. In particular, the so-called scientific purpose CMOS sensors have noise plus dark current as small as a few electrons even at room temperature. It means that those sensors can in principle detect x-rays in photon-counting mode. In fact, such CMOS sensors are employed in the rocket experiment, FOXSI3 to observe solar x-rays\cite{Ishikawa2018}, and are planned to be employed in the x-ray wide-field survey mission Einstein Probe\cite{Wang2018}. In this paper, we have employed a newly released 2.5-$\mathrm{\mu}$m pixel CMOS sensor designed for visible light to measure its x-ray spectroscopic and polarimetric performance. This is the smallest pixel size detector whose x-ray performance has been measured so far. When errors are quoted without further specification, these refer to 1$\mathrm{\sigma}$ uncertainties in this paper.

\section{Detection of X-ray Events with GMAX0505}
\subsection{CMOS Image Sensor, GMAX0505}

\begin{figure} [H]
   \begin{minipage}{0.5\hsize}
   \begin{center}
   \begin{tabular}{c} 
 \includegraphics[width=7cm]{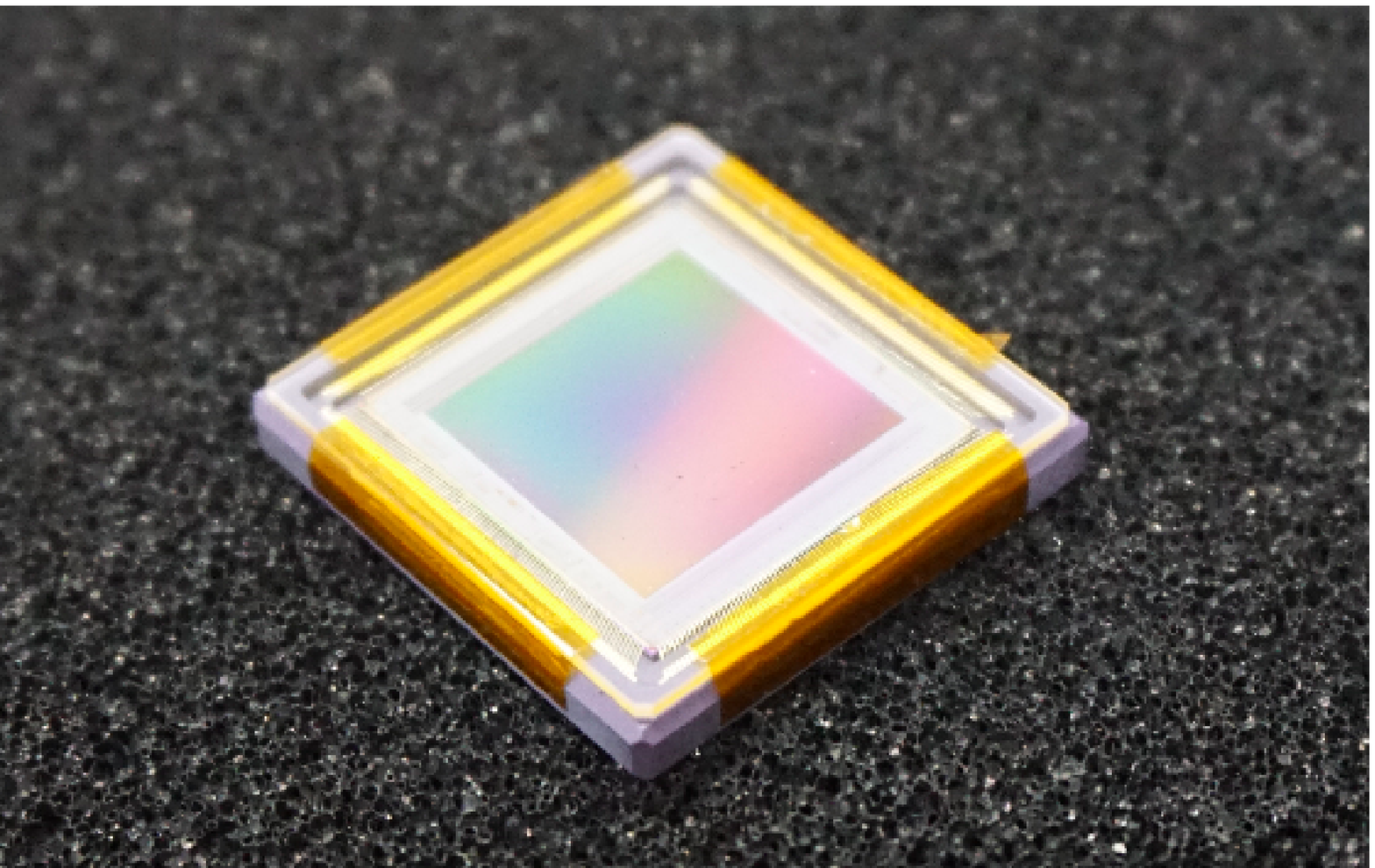}
   \end{tabular}
   \end{center}
   \caption[GMAX0505 image]{Image of the CMOS sensor of GMAX0505.}
   \label {GMAX_image}
 \end{minipage}
\begin{minipage}{0.5\hsize}
\begin{table}[H]
\begin{center}
\caption[GMAX0505 properties] {GMAX0505 properties.}
 \label {GMAX_table}
\vspace{0.2cm}
  \begin{tabular}{@{} lc @{}} \hline 
chip size  & $15.85~\mathrm{mm} \times 16.88~\mathrm{mm}$ \\
pixel size & $2.5~\mu\mathrm{m} \times 2.5~\mu\mathrm{m}$ \\
number of pixels    & $5120 \times5120$ \\
Effective Area & $12.8~\mathrm{mm}\times12.8~\mathrm{mm}$ \\
Frame rate & 40 frames per second @12 bit \\
Shutter Type & Global shutter\\
Device Type & Front Illuminated \\ \hline
  \end{tabular}
  \end{center}
\end{table}
\end{minipage}
\end{figure}

GMAX0505 is a CMOS image sensor developed by GPixel Inc. and is originally designed for visible light imaging, displayed in Fig \ref{GMAX_image}. Its imaging area consists of  $5120\times5120$ pixels, and the size of a pixel is 2.5\,$\mathrm{\mu}$m $\times$ 2.5 $\mathrm{\mu}$m, which can provide us high spatial resolution. Table \ref{GMAX_table} shows the specific properties of GMAX0505. The structure of GMAX0505 is designed for visible light. In particular, its advanced light pipe structure optimizes the response to visible light as described in Yokoyama et al.\cite{Yokoyama2018} We applied it for x-ray imaging and polarimetry for the first time. Although Yokoyama et al.\cite{Yokoyama2018} provide a schematic view of the pixel structure, the size of the photodiode, which is essential for the x-ray detection, is not provided. We thus estimate the thickness of the x-ray detection layer to be $\sim$5\,$\mathrm{\mu}$m in Sec. 3.3. Although the device is usually equipped with a cover glass, we employ the device without it. However, micro-lenses implemented in the sensor at the illumination side are not removed in our experiment. As this sensor is very sensitive to the visible light, we cover the sensor with a dark curtain to block the visible light.

We adopt the evaluation board and software developed by GPixel Inc. to operate GMAX0505 and acquire data. GMAX0505 has 32 levels of gain, and we utilize two of them; when a register is set to 0, GMAX0505 is operated in a low-gain mode. While the register is set to 4, it is in a high-gain mode. GMAX0505 is operated at room temperature and the atmospheric condition throughout our experiments in this paper. Readout noise is evaluated from data with an exposure time of 1\,ms. The average noise is 5.7\,e${}^{-}$ (RMS) in the low-gain mode and 2.7\,e${}^{-}$ (RMS) in the high-gain mode.

\subsection{Detecting X-rays from ${}^{55}$Fe}

First, we used ${}^{55}$Fe as an x-ray source to see if GMAX0505 can detect x-ray photons. Both high-gain and low-gain modes were adopted in the ${}^{55}$Fe measurement in order to check properties with these gains. We took 500 frames in each of the high- and low-gain modes with the exposure time of 100\,ms. About 200 frames were also obtained with the same exposure without the x-ray source as background data. We estimated the background level of each pixel by averaging all the frames in the background data and subtracted them from the pixel level of each pixel of each frame with the x-ray illumination.

\begin{figure} [H]
       \begin{center}
     \begin{tabular}{c} 
       \includegraphics[clip,scale=0.45]{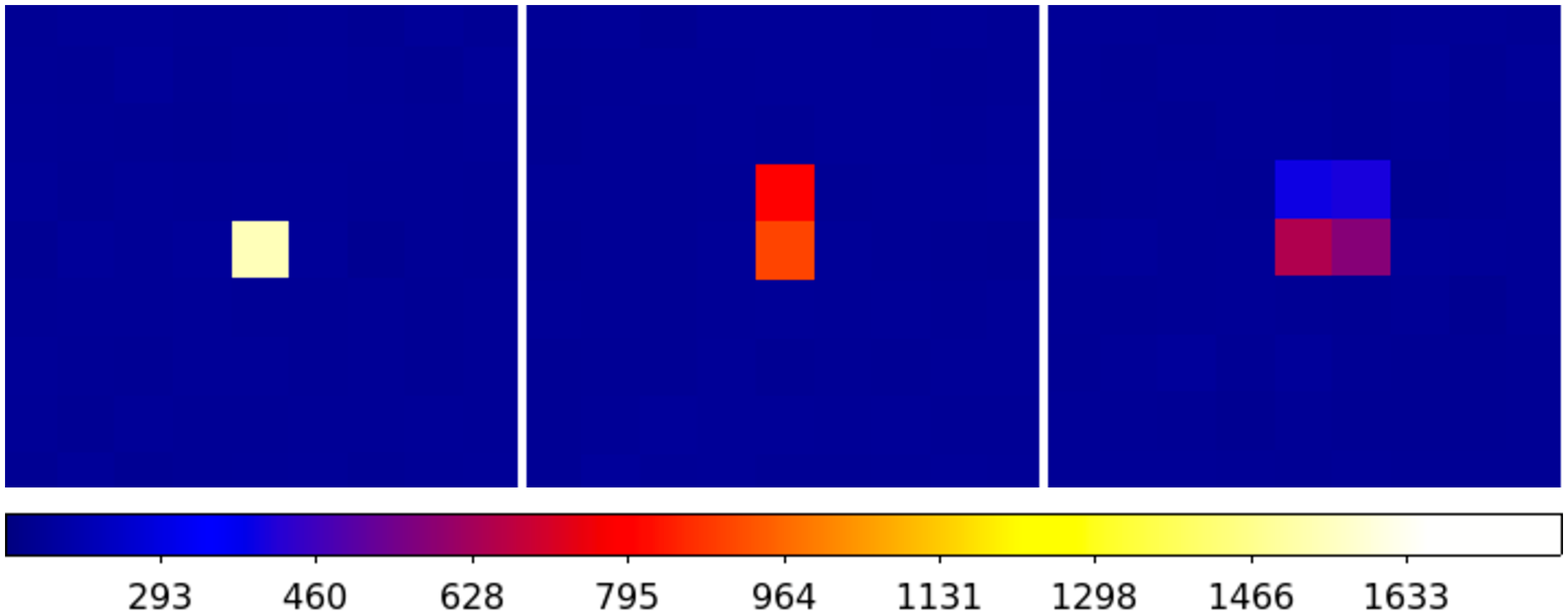}
     \end{tabular}
   \end{center}
  \caption[Frame images of GMAX0505 exposed to x-rays from ${}^{55}$Fe ]{Zoom-in frame images of a single-pixel event (left), a double-pixel event (middle), and an extended event (right) obtained by GMAX0505 in the high-gain mode exposed to x-rays from ${}^{55}$Fe. More than 60\% of events are single-pixel events.}
    \label{Fe_image}
  \end{figure}

Figure \ref{Fe_image} shows a zoom-in frame data obtained in the ${}^{55}$Fe irradiation test. We clearly detect x-ray events. Most of the events are confined within one pixel, which we call single-pixel events. However, we found some events spread over neighboring multiple pixels, which are often seen in x-ray detection with x-ray CCDs. If the event spreads over two adjacent pixels, we call it double-pixel events. Events spreading over more than two pixels are defined as extended events. According to the standard event extraction scheme employed for x-ray CCDs, we employ two signal thresholds for event extraction, the event threshold for the event center, and the split threshold for neighboring pixels. We set the event threshold to 10$\sigma$ and the split threshold to 3$\sigma$, where $\sigma$ is the standard deviation of the background level. In our analysis, we classify the events into single-pixel, double-pixel, and extended events.

\begin{figure} [H]
      \begin{center}
    \begin{tabular}{c} 
      \includegraphics[clip,scale=0.45]{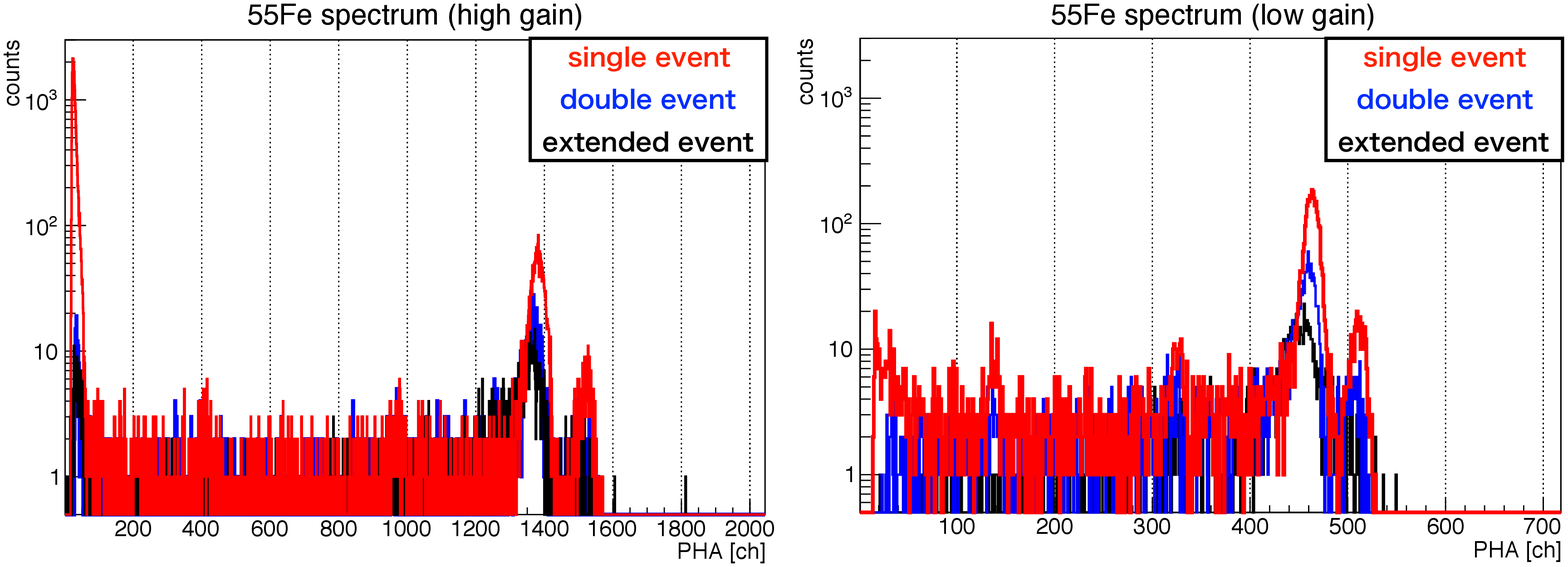}
    \end{tabular}
  \end{center}
 \caption[Spectra obtained by GMAX0505 exposed to x-rays from ${}^{55}$Fe]{(Left) A spectrum obtained by GMAX0505 exposed to x-rays from ${}^{55}$Fe in the high-gain mode. The peaks around 1400 and 1500\,ch correspond to 5.9\,keV (Mn K$\mathrm{\alpha}$) and 6.4\,keV (Mn K$\mathrm{\beta}$), respectively. (Right) Same as left, but in the low-gain mode.}
   \label{Fe_spec}
 \end{figure}

 Figure \ref{Fe_spec} shows the spectra of single-pixel, double-pixel, and extended events obtained by GMAX0505 exposed to x-rays from ${}^{55}$Fe. It indicates that GMAX0505 evidently detected x-rays from the source. The gain of the high-gain mode is calculated to be 4.27\,eV ch${}^{-1}$, whereas that of the low-gain mode is 12.7\,eV ch${}^{-1}$ for single-pixel events. The energy resolution is also obtained from the width of the Mn K$\mathrm{\alpha}$ peak; FWHM at 5.9\,keV is 176\,eV in the high-gain mode and 196\,eV in the low-gain mode.


\section{ X-ray Beamline Experiment}

\subsection{SPring-8 BL20B2 and Calibration of the Beam Polarization}

To evaluate the x-ray detection efficiency and the x-ray polarimetry sensitivity of GMAX0505, we conducted an x-ray beam experiment at BL20B2 in SPring-8, the synchrotron radiation facility in Hyogo, Japan in October 2018. The beamline was 215\,m in length, and the beam divergence was as small as sub-arcsecond. We fixed our CMOS detector in the experimental hatch at the beam downstream. The beam size was collimated by a slit at the entrance of the hatch to 10 mm $\times$ 10 mm. X-ray photons were monochromatized by a double-crystal monochromator set at upstream of the beam. We used the x-ray energy either 12.4 or 24.8\,keV and inserted attenuators so that we can avoid the event pile-up; a 0.15\,mm-thick Mo plate to 12.4-keV beam, or a 0.1\,mm-thick Sn plus 0.1\,mm-thick Cu plate to 24.8-keV beam.

We prepared a scattering polarimeter system consisting of a Be target of 5-mm diameter and 20-mm length and a CdTe detector, XR-100CdTe provided by AMPTEK, Inc. to detect x-ray photons scattered by 90 deg. We applied the x-ray beam to this system for the measurement of the x-ray beam polarization. The modulation of the count rate was obtained to be 94.05\% $\pm$ 0.03\% at 12.4\,keV and 93.26\% $\pm$ 0.08\% at 24.8\,keV. We calculated MF of our scattering polarimeter system to be 94.7\%, simply from its geometry and the differential cross section of the Thomson scattering. Dividing the former by the latter, we obtained the polarization degree of the beam to be 99.31\% $\pm$ 0.03\% at 12.4\,keV and 98.48\% $\pm$ 0.08\% at 24.8\,keV, respectively. The polarization direction (electric vector) is parallel to the horizontal plane.

\subsection{X-ray Beam Irradiation to GMAX0505}

\begin{figure} [H]
   \begin{center}
   \begin{tabular}{c} 
 \includegraphics[clip,scale=0.45]{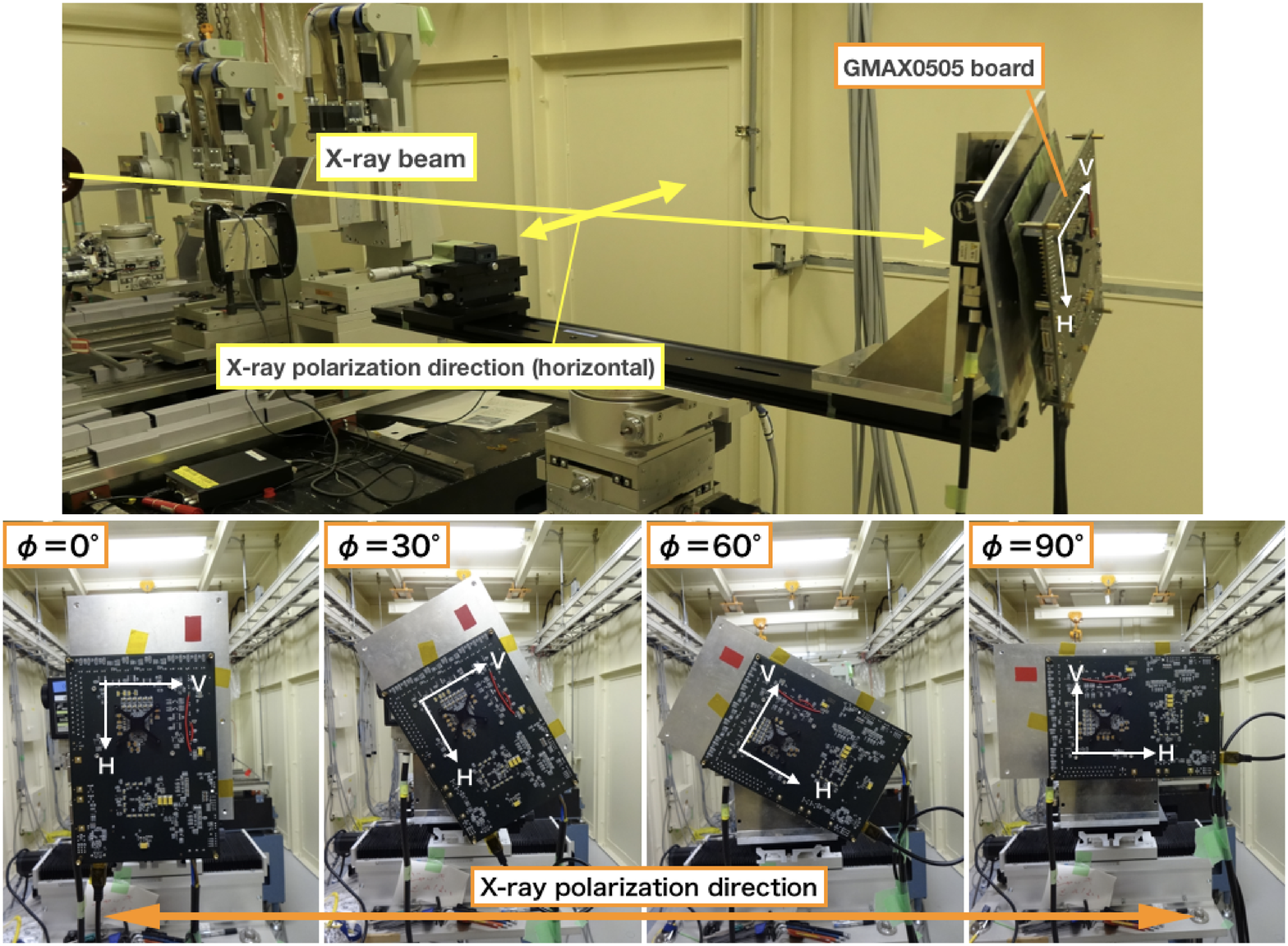}
   \end{tabular}
   \end{center}
   \caption[Experimental Setup]{Experimental setup in SPring-8 BL20B2. Distance between GMAX0505 and the beamline window are set to $\sim$ 2.5\,m. The imaging area of GMAX0505 has H and V axis, and hence we define the rotation angle $\phi$ as legends in the figure.}
   \label {setup}
   \end{figure}

The experimental setup is shown in Fig. \ref{setup}. We defined the H and V direction in the GMAX0505 imaging area and the rotation angle $\phi$ of GMAX0505 as shown in Fig. \ref{setup}. We took frame data at $\phi$ = 0 deg, 30 deg, 60 deg, and 90 deg. The number of obtained frames was 200 to 500 at each angle and each energy, and an exposure of every frame was 600\,ms at 12.4\,keV, and 5\,ms at 24.8\,keV. We adopted the high-gain mode when we measured at 12.4\,keV, and the low-gain mode at 24.8\,keV. Background data were also obtained with the same exposure without irradiating the x-ray beam.

\begin{figure} [H]
        \begin{center}
      \begin{tabular}{c} 
        \includegraphics[clip,scale=0.45]{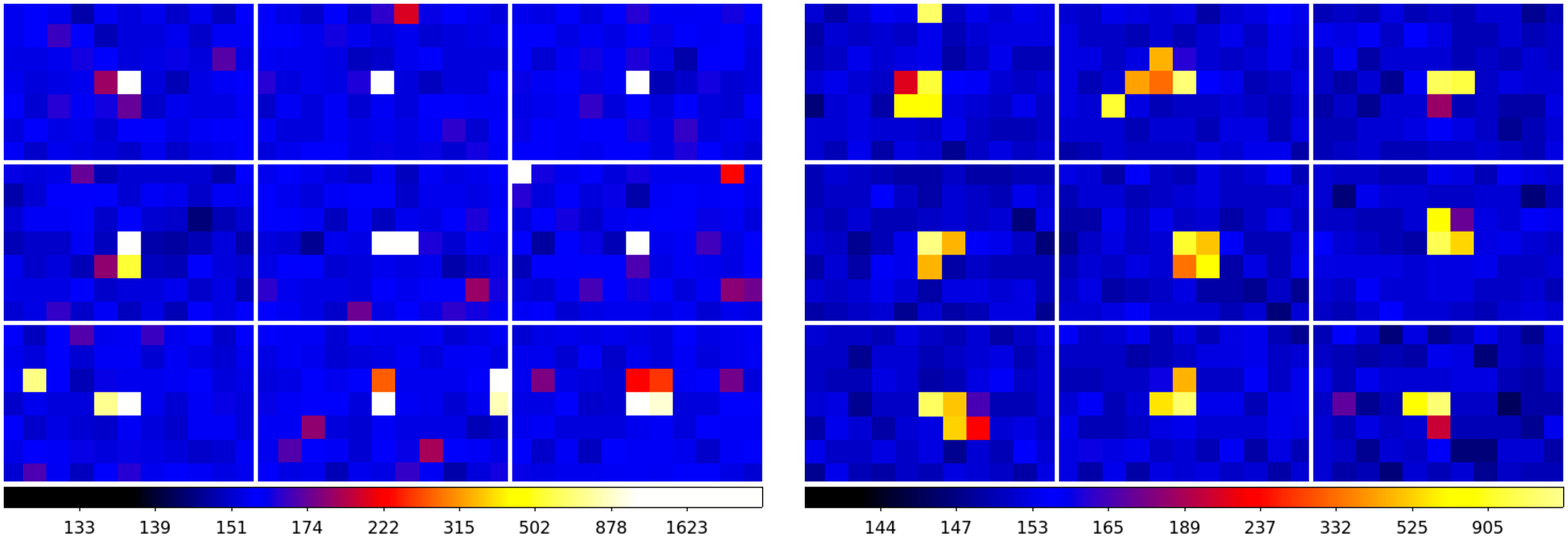}
      \end{tabular}
    \end{center}
   \caption[Zoom-in frame images of GMAX0505 exposed to the x-ray beam]{(Left) Zoom-in frame images obtained by GMAX0505 exposed to the x-ray beam at 12.4\,keV. These events are picked up at random in a frame data. (Right) Same as left, but the energy is 24.8\,keV. Almost all events split into neighboring pixels, whereas some events are confined in one pixel at 12.4\,keV. }
     \label{beam_image}
   \end{figure}

\begin{figure} [H]
        \begin{center}
        \begin{tabular}{c} 
        \includegraphics[clip,scale=0.4]{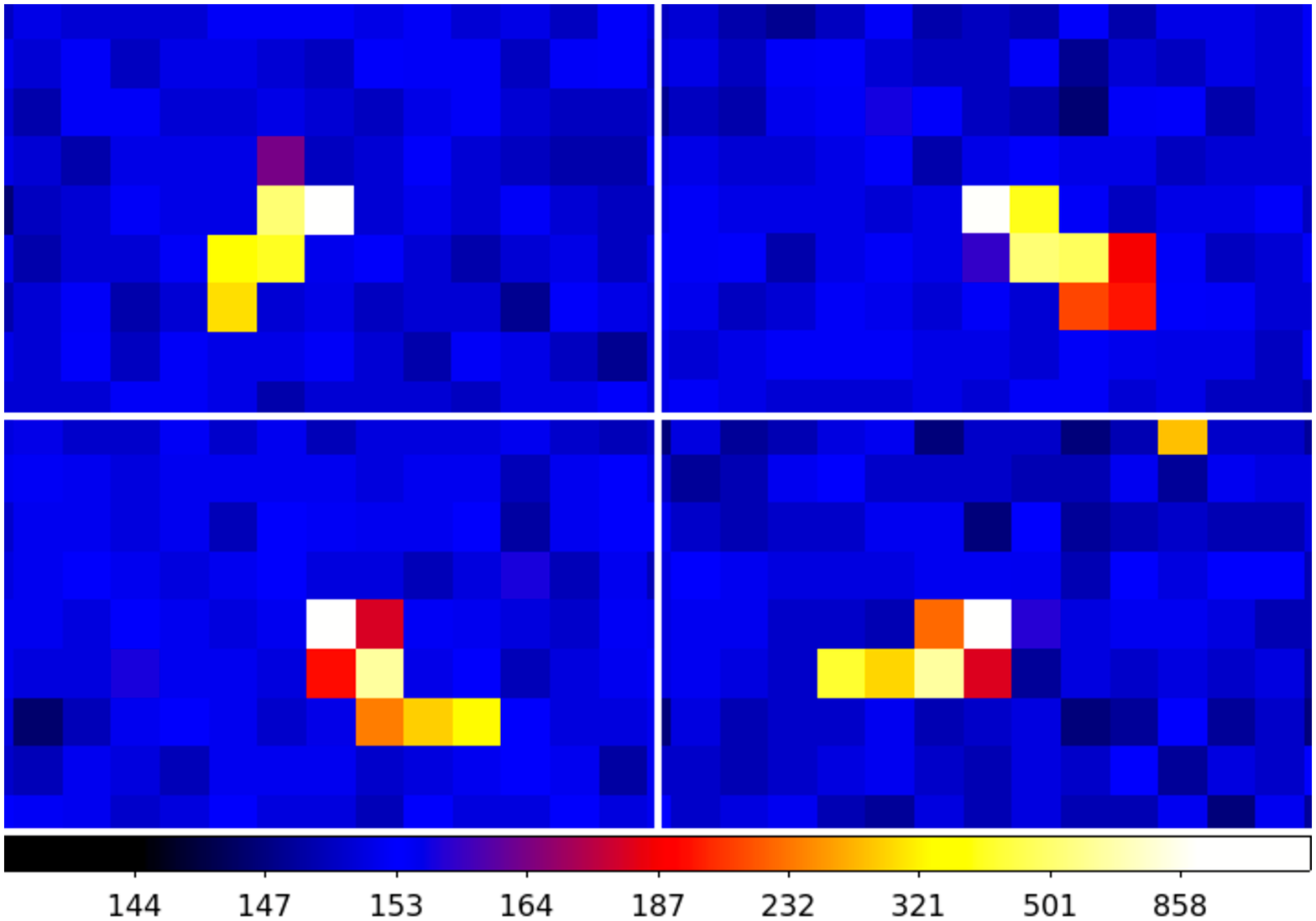}
        \end{tabular}
        \end{center}
        \caption[Zoom-in frame images of photoelectron tracks]{Zoom-in frame images of extended events at 24.8\,keV. We extract the events where the track of a photoelectron is especially visible.}
       \label{beam_image2}
     \end{figure}

We show x-ray events detected in the raw frame data of GMAX0505 in Fig. \ref{beam_image}. The numbers of single-pixel, double-pixel, and extended events are almost the same at 12.4\,keV, whereas most of the 24.8-keV events are extended over multiple pixels. This contrast is caused by the difference in the range of photoelectrons in silicon, 1.1\,$\mathrm{\mu}$m for the 12.4\,keV x-ray incidence and 4.3\,$\mathrm{\mu}$m for the 24.8\,keV x-ray incidence, according to the empirical formula of $r~(\mathrm{{\mu}m})=\left[E_\mathrm{e}/10~(\mathrm{keV})\right]^{1.75}$, where $E_\mathrm{e}$ is the initial photoelectron energy\cite{Janesick1985}. The small pixel size of 2.5\,$\mathrm{\mu}$m for GMAX0505 enables us to "image" photoelectron tracks at least for the 24.8\,keV x-ray incidence as shown in Fig. \ref{beam_image2}. To our knowledge, this is the first time for solid-state detectors to take such images.

\begin{figure} [H]
     \begin{center}
       \begin{tabular}{c} 
         \includegraphics[clip,scale=0.45]{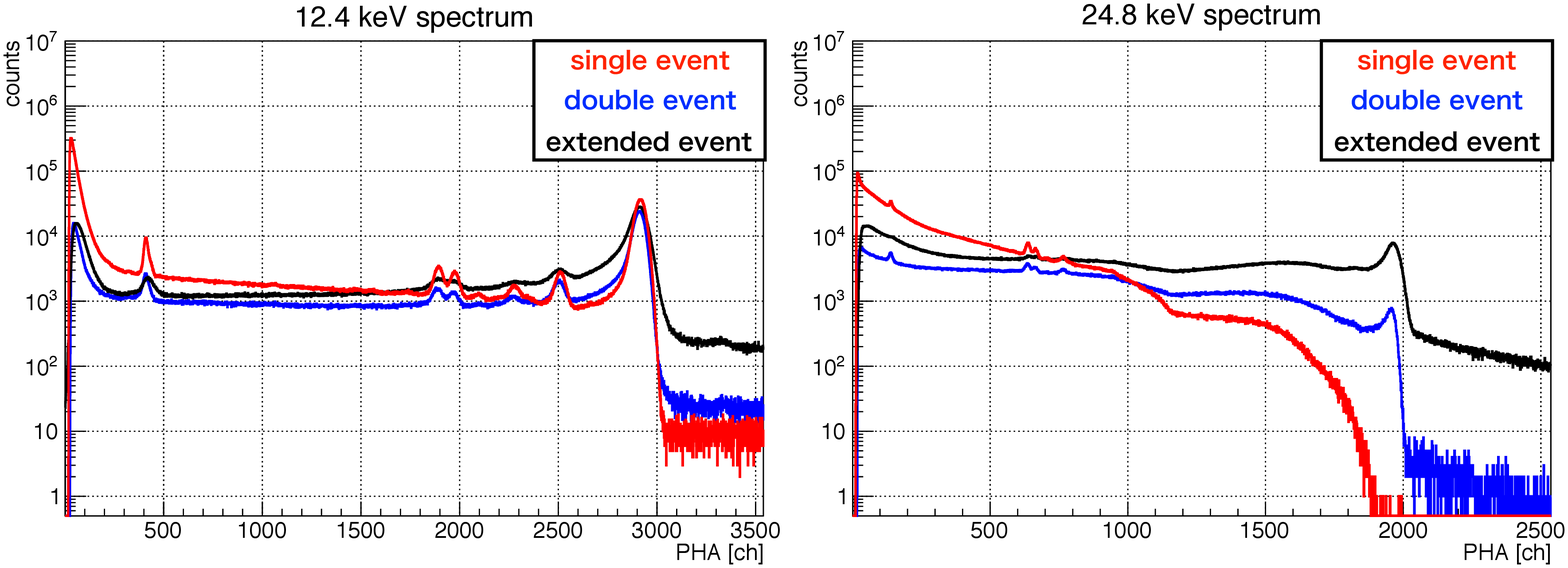}
        \end{tabular}
      \end{center}
      \caption[Spectra of GMAX0505 exposed to the x-ray beam]{(Left) Spectra of GMAX0505 exposed to the 12.4-keV x-ray beam. Different colors show different types of events. We utilized a range of 2600 to 3200\,ch to extract events of which energy is around 12.4\,keV. (Right) Same as left, but the beam energy is 24.8\,keV. We utilize the events in 1900 to 2100\,ch.}
  \label{beam_spectra}
\end{figure}

The spectra measured with GMAX0505 are shown in Fig. \ref{beam_spectra}. Subtraction of background data and event selection are conducted in the same way as described in Sec. 2.2. Each spectrum has a primary peak corresponding to the incident x-ray beam energy 12.4 or 24.8\,keV. Other peaks are escape events and fluorescence x-ray events from surrounding materials.

\subsection{X-ray Detection Efficiency of GMAX0505}

To measure the detection efficiency of GMAX0505 to x-rays, we also measured the 12.4\,keV x-ray beam by the CdTe detector that was used in Sec. 3.1. Exposures of the CdTe observations were set to 600\,s. Attenuation plates and distance between the beamline window and the detector were set to same as those in the measurement with GMAX0505. The CdTe detector had a CdTe crystal with thickness of 1\,mm, and its detection efficiency was more than 99.9\% at 12.4-keV x-rays. Hence, we regard that its detection efficiency was 100\%, and compared count rates per 1\,mm$^2$ derived by GMAX0505 and the CdTe sensor. The detection efficiency was calculated to be 1.9\% at 12.4\,keV, from which the thickness of the detection layer was evaluated to be $\sim$5\,$\mathrm{\mu}$m. Note that we simplified the structure of the detection layer to be a flat slab of Si in our estimation, and it was not necessarily the physical size of the photodiode. In this calculation, we integrated all types of x-ray events in 600 to 3200\,ch. For the 24.8\,keV x-ray incidence, the integrated counts were mostly in the low-energy tail of the spectra, and it was difficult to evaluate it accurately. However, we employed only the double-pixel events to detect the polarization of x-rays, as discussed in detail in Sec. 4. When we employ the events in the pulse height range specifically described in Fig. \ref{beam_spectra}, we obtained the detection efficiency of double-pixel events at 12.4 and 24.8\,keV to be 0.093\% and 0.0011\%, respectively.

\section{X-ray Polarimetry with GMAX0505}

\subsection{Polarization Measurement with Double-pixel Events}

Photoelectrons are preferentially emitted in the direction parallel to the electric vector of incident x-ray photons, i.e., x-ray events should spread horizontally if the beam is polarized horizontally. X-ray CCD image sensors show the same characteristics and can detect x-ray polarization by selecting double-pixel events\cite{Tsunemi1992}. We then focus on the double-pixel events and estimate the polarimetry sensitivity of our GMAX0505.

\begin{figure} [hb]
   \begin{center}
   \begin{tabular}{c} 
 \includegraphics[clip,scale=0.35]{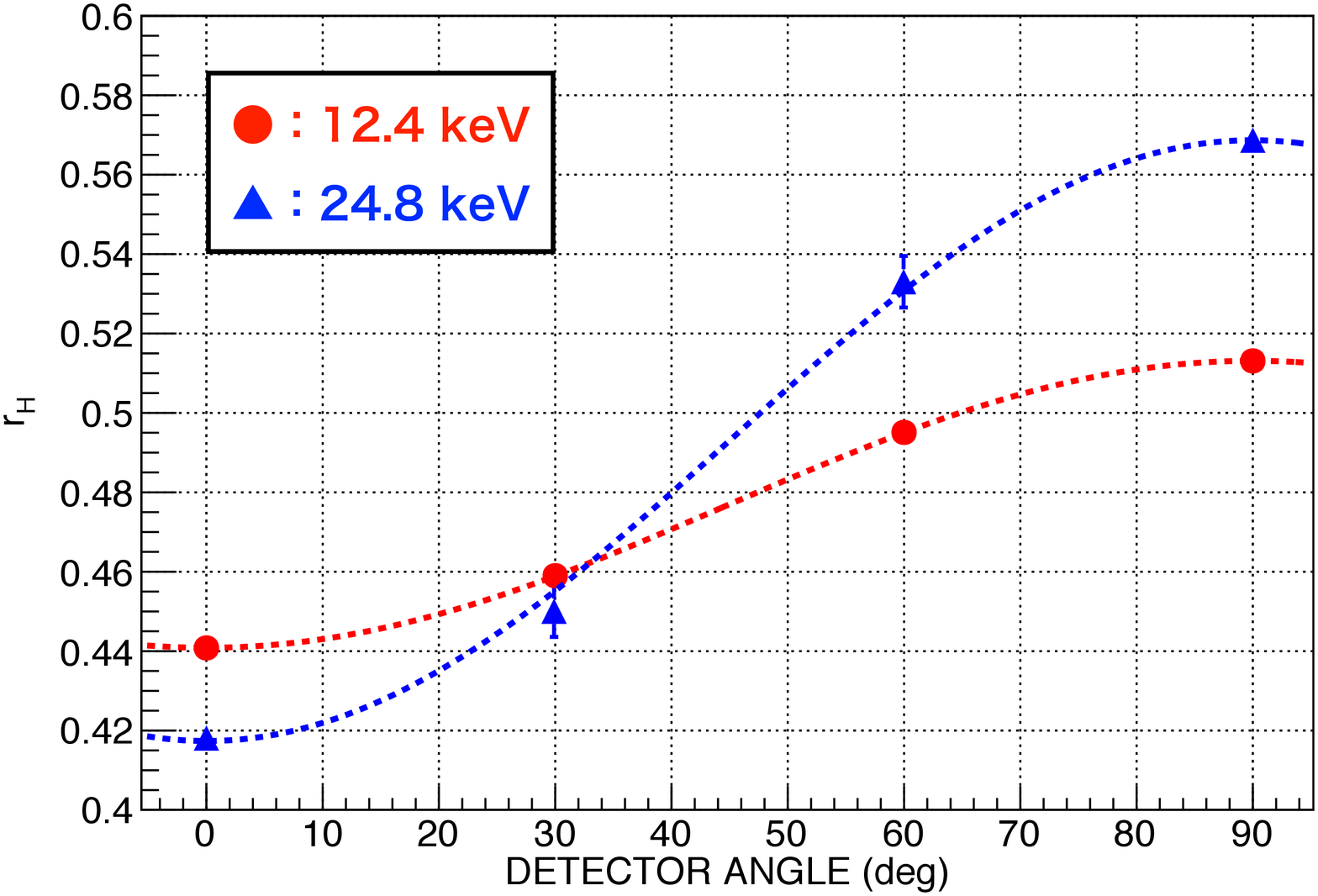}
   \end{tabular}
 \end{center}
\caption[$r_\mathrm{H}(\phi)$ at 12.4\,keV and 24.8\,keV]{The number ratio of double events spreading along the H axis against the whole double events defined as $r_\mathrm{H}(\phi)$ in Eq. (\ref{eq2}). Red shows $r_\mathrm{H}(\phi)$ at 12.4\,keV, and blue shows $r_\mathrm{H}(\phi)$ at 24.8\,keV. Errors of 12.4\,keV $r_\mathrm{H}$ are negligible because of sufficient statistics. These plots can be well fitted by a sine curve function that a modulation curve is expected to follow.}
  \label{ratio}
   \end{figure}

We calculate the number ratio of double-pixel events spreading along the H axis against the total double-pixel events in every rotation angle by the following equation:
\begin{align}
  \label{eq2}
  r_\mathrm{H}(\phi)=\frac{N_\mathrm{H}(\phi)}{N_\mathrm{H}(\phi)+N_\mathrm{V}(\phi)} ,
\end{align}
where $\phi$ shows the rotation angle of GMAX0505, and $N_\mathrm{H}(\phi)$ and $N_\mathrm{V}(\phi)$ represent the number of double-pixel events of GMAX0505 along its H and V axes. We adopt the spectral pulse height range of 2600 to 3200\,ch and 1900 to 2100\,ch at 12.4 and 24.8\,keV, respectively. When $\phi$ gets closer to 90 deg, i.e., when the GMAX0505 H axis gets closer to the beam polarization direction, $r_\mathrm{H}$ gradually increases as shown in Fig. \ref{ratio}, as we anticipate. It indicates that we succeed in detecting x-ray polarization with GMAX0505. We also calculate $r_\mathrm{H}$ in case the incident x-rays are nonpolarized using the double-pixel events of ${}^{55}$Fe data described in Sec. 2.2. We obtain $r_\mathrm{H}$ of 46.6\% $\pm$ 2.6\%, which shows that carriers may be likely to spread along the V axis in GMAX0505 even if incident x-rays are nonpolarized.

To evaluate the polarimetry sensitivity, we calculate the MF, which is defined as
\begin{align}
  \label{eq3}
  MF=\frac{1}{P}\frac{r_\mathrm{H}(90~\mathrm{deg})-r_\mathrm{H}(0~\mathrm{deg})}{r_\mathrm{H}(90~\mathrm{deg})+r_\mathrm{H}(0~\mathrm{deg})} ,
\end{align}
where P describes the degree of polarization of the incident x-ray beam, and we substitute the value derived in Sec. 3.1 to it. In consequence, we obtain a MF of 7.63\% $\pm$ 0.07\% at 12.4\,keV, and 15.5\% $\pm$ 0.4\% at 24.8\,keV.

\subsection{Comparison between Simulations and Measurements}

We compare the $r_\mathrm{H}$ values obtained by GMAX0505 with numerical simulations performed by the Geant4 software\cite{Geant2003, Geant2006, Geant2016} in an ideal case where diffusion of carriers in silicon is neglected, and electric noise is zero. We choose the Geant4 10.03.03 version and adopt Livermore model as a physics model. This model includes fluorescence x-rays, Auger electrons, and polarization effect. In our simulations, we trace electrons (not only the primary photoelectrons but also the secondary electrons generated by various processes) until its energy is reduced to 10\,eV. We construct 5$\times$5 Si pixel arrays, where each pixel size is the same as that of GMAX0505. The thickness of pixels is estimated to be 5\,$\mathrm{\mu}$m in Sec. 3.3, and here we adopt this value. In the simulations, $10^8$ x-ray photons with 100\% polarization enter the pixels. We calculate the total energy deposit at each pixel for each event and compare the numbers of double-pixel events parallel and perpendicular to the polarization direction. MF obtained from this simulation is 21.4\% $\pm$ 0.6\% at 12.4\,keV, and 20.0\% $\pm$ 0.3\% at 24.8\,keV. These values are larger than the measured values by a factor of 1.3 to 3. It would be attributed to our simplified assumptions; for example, the weak electric field in Si substrate between and under the photodiodes may affect the signal of double-pixel events. These values should be regarded as theoretical upper limits of MF in the ideal case.

\subsection{Discussion}

The previous polarimetry measurement by a 12-$\mathrm{\mu}$m pixel size CCD detector has been conducted at 27 and 43\,keV.\cite{Hayashida1999} Since the definition of their MF is different from ours, we convert MF to be 7\% at 27\,keV and 15\% at 43\,keV with Eq. (\ref{eq3}). At this time, we perform polarimetry in the lower energy band with the higher spatial resolution than that in the CCD experiment, owing to the smaller pixel size of GMAX0505. It is also remarkable that GMAX0505 is operated at room temperature, in contrast to x-ray CCDs, which need to be cooled down to about $-$100${}^{\circ}$C. Another advantage of the CMOS over CCDs is pile-up tolerance. In fact, we need to attenuate x-rays by orders of magnitude if we employ a CCD with a frame time of a few seconds.

Here, we evaluate the sensitivity of the x-ray polarimetry as minimum detectable polarization (MDP). MDP is calculated as
\begin{align}
  \label{eq4}
  \mathrm{MDP}=\frac{4.29}{MF \times \sqrt{N_{\mathrm{H}}+N_{\mathrm{V}}} } ,
\end{align}
where $N_\mathrm{H}+N_\mathrm{V}$ is the total number of double-pixel events\cite{Weisskopf2010}. MDP shows minimum polarization degree required to detect x-ray polarization at the 99\% confidence level. We calculate MDP in the case that we observe the Crab Nebula in the 10 to 20\,keV band with GMAX0505. The spectrum of the Crab Nebula is approximated by a power-law continuum with the photon index of 2.12 and the normalization at 1\,keV of 9.42\,photons $\mathrm{keV}^{-1}~\mathrm{cm}^{-2}~\mathrm{s}^{-1}$\cite{Kirsch2005}. With the power-law emission with this model, the photon flux in 10 to 20\,keV is 0.344\,$\mathrm{photons}~\mathrm{cm}^{-2}~\mathrm{s}^{-1}$. If an exposure time is $10^6$\,s and the effective area of a mirror is 400\,$\mathrm{cm}^2$, $1.38\times10^8$ x-ray photons enter the detector. Since the detection efficiency of double-pixel events is 0.093\% $\pm$ 0.002\% at 12.4\,keV in Sec. 3.3, we substitute 1.28$\times10^5$ to $N_\mathrm{H}+N_\mathrm{V}$. Adopting MF = 7.63\% $\pm$ 0.07\% at 12.4\,keV, we obtain MDP of $\sim$16\% in the 10 to 20\,keV band.

Note that our data reduction procedure has not yet been optimized to the polarimetry with GMAX0505. We employ only double-pixel events. As shown in Fig. \ref{beam_image}, particularly for the 24.8-keV incidence, photoelectron track images extending over multiple pixels should have valid information on their emission direction. Extensive studies have been made to analyze photoelectron track images obtained with gas micro-pixel detectors for x-ray polarimetry. We expect that those techniques are also useful to those in GMAX0505, and they will improve MDP significantly.

MF of GMAX0505 is not as high as other x-ray polarimeters; 20\% to 65\% for the gas photoelectron tracking polarimeter at 2 to 8\,keV\cite{Weisskopf2016} and 30\% to 50\% for the hard x-ray and soft $\gamma$-ray scattering polarimeters above 20\,keV.\cite{Yonetoku2011, Hayashida2016b, Chauvin2017} However, the energy range of 10 to 20\,keV is the valley of these two classes of x-ray polarimeters, and there is a room for the small-sized pixel CMOS sensors such as GMAX0505. Combining with a future high-quality x-ray mirror of 10-m focal lengths, we would obtain the spatial resolution of $\sim$0.05\,arcsec. This will enable us to detect spatially resolved scattering x-rays from molecular tori surrounding supermassive black holes in nearby galactic nuclei, for example. This type of small pixel CMOS sensors is also essential for the newly proposed x-ray interferometer without mirrors, Multi-Image X-ray Interferometer Module.\cite{Hayashida2016, Hayashida2018}

\section{Conclusion}
\label{sec:sections}

We applied GMAX0505, the CMOS image sensor designed for visible light and that has the smallest pixel size ever, 2.5\,$\mathrm{\mu}$m $\times$ 2.5 $\mathrm{\mu}$m each, to x-ray imaging, spectroscopy, and polarimetry. We first irradiated x-rays from ${}^{55}$Fe and obtained the energy resolution of 176\,eV (FWHM) at 5.9\,keV at room temperature operation. We then brought GMAX0505 to SPring-8 BL20B2 and measured its x-ray polarimetry sensitivity. We obtained MF of 7.63\% $\pm$ 0.07\% at 12.4\,keV and 15.5\% $\pm$ 0.4\% at 24.8\,keV. These results show that GMAX0505 has both x-ray polarimetry sensitivity and spectroscopic performance with the highest spatial resolution ever.


\acknowledgments
The synchrotron radiation experiments were performed at BL20B2 in SPring-8 with the approval of the Japan Synchrotron Radiation Research Institute (JASRI) (Proposals Nos. 2018B1235, 2018A1368, 2017B1186, and 2017B1098). KH acknowledges the support from JSPS KAKENHI under Grant Nos. JP18K18767, JP16K13787, JP16H00949, and JP26109506. HM acknowledges the support from JSPS KAKENHI under Grant No. 15H02070. HA acknowledges the support from JSPS KAKENHI under Grant Nos. 15H02070 and 17K18782. HN acknowledges the support from JSPS KAKENHI under Grant Nos. 15H03641 and 18H01256. The authors have no relevant financial interests in the manuscript and no other potential conflicts of interest to disclose.


\bibliography{GMAX0505_paper}   
\bibliographystyle{spiejour}   

\vspace{2ex}\noindent\textbf{Kazunori Asakura} is a graduate student at Osaka University in Japan, He received his BS degree in Physics from Osaka University in 2018. His current research includes instrumentation of X-ray imaging spectroscopy and polarimetry (especially CCDs and CMOS), and observational study using archival data of X-ray astronomy satellites.


\listoffigures
\listoftables

\end{document}